
%
%
%
%

\message{*** SIAM Plain TeX Proceedings Series macro package, version
1.0,
November 6, 1992.***}

\catcode`\@=11

\baselineskip=14truept


\hsize=36truepc
\vsize=55truepc
\parindent=18truept
\def\firstpar{\parindent=0pt\global\everypar{\parindent=18truept}}
\parskip=0pt


\font\tenrm=cmr10
\font\tenbf=cmbx10
\font\tenit=cmti10
\font\tensmc=cmcsc10
\def\tenpoint{%
   \def\rm{\tenrm}\def\bf{\tenbf}%
   \def\it{\tenit}\def\smc{\tensmc}
        \textfont0=\tenrm \scriptfont0=\sevenrm
	\textfont1=\teni \scriptfont1=\seveni
	\textfont2=\tensy \scriptfont2=\sevensy
	\textfont3=\tenex \scriptfont3=\tenex
\baselineskip=12pt\rm}%

\font\ninerm=cmr9
\font\ninebf=cmbx9
\font\nineit=cmti9
\def\ninepoint{%
   \def\rm{\ninerm}\def\bf{\ninebf}%
   \def\it{\nineit}\baselineskip=11pt\rm}%

\fontdimen13\tensy=2.6pt
\fontdimen14\tensy=2.6pt
\fontdimen15\tensy=2.6pt
\fontdimen16\tensy=1.2pt
\fontdimen17\tensy=1.2pt
\fontdimen18\tensy=1.2pt

\font\ninerm=cmr9
\font\elevenrm=cmr10 scaled\magstephalf
\font\fourteenrm=cmr10 scaled\magstep 1
\font\eighteenrm=cmr10 scaled\magstep 3
\font\twelvebf=cmbx10 scaled\magstep 1
\font\elevenbf=cmbx10 scaled\magstephalf

\def\textfont{\elevenrm}

\def\headfont{\twelvebf}
\def\smallheadfont{\elevenbf}
\def\titlefont{\eighteenrm}

\def\authorfont{\fourteenrm}
\def\rheadfont{\tenrm}
\def\abstractfont{\tenrm}
\def\smc{\tensmc}

\def\footnote#1{
\baselineskip=11truept\edef\@sf{\spacefactor\the\spacefactor}#1\@sf
  \insert\footins\bgroup\ninepoint\hsize=36pc
  \interlinepenalty10000 \let\par=\endgraf
   \leftskip=0pt \rightskip=0pt
   \splittopskip=10pt plus 1pt minus 1pt \floatingpenalty=20000
\smallskip
\item{#1}\bgroup\baselineskip=10pt\strut
\aftergroup\@foot\let\next}
\skip\footins=12pt plus 2pt minus 4pt
\dimen\footins=36pc


\def\startchapter{\topinsert\vglue54pt\endinsert}

\def\title#1\endtitle{\titlefont\centerline{#1}\vglue5pt}
\def\lasttitle#1\endlasttitle{\titlefont\centerline{#1}\vskip1.32truepc}
\def\author#1\endauthor{\authorfont\centerline{#1}\vglue8pt\textfont}
\def\lastauthor#1\endlastauthor{\authorfont\centerline{#1}
\vglue2.56pc\textfont}
\def\abstract#1\endabstract{\baselineskip=12pt\leftskip=2.25pc
     \rightskip=2.25pc\abstractfont{#1}\textfont}


\newcount\headcount
\headcount=1
\newcount\seccount
\seccount=1
\newcount\subseccount
\subseccount=1
\def\secreset{\global\seccount=1}
 \def\subsecreset{\global\subseccount=1}


\def\headone#1{\baselineskip=14pt\leftskip=0pt\rightskip=0pt
\vskip17truept\parindent=0pt
{\headfont\the\headcount\hskip14truept #1}
\par\nobreak\firstpar\global\advance\headcount by 0
   \global\advance\headcount by 1\secreset\vskip2truept\textfont}

\def\headtwo#1{\advance\headcount by -1%
   \vskip17truept\parindent=0pt{\headfont\the\headcount.%
   \the\seccount\hskip14truept #1}
   \global\advance\headcount by 1\global\advance\seccount by 1
   \global\advance\subseccount by 1\subsecreset\vskip2pt\textfont}

 \def\headthree#1{\advance\headcount by -1\advance\seccount by -1
   \vskip17truept\parindent=0pt{\smallheadfont\the\headcount.%
   \the\seccount.\the\subseccount\hskip11truept #1}\hskip6pt\ignorespaces
   \firstpar\global\advance\headcount by 1\global\advance\seccount by 1
   \global\advance\subseccount by 1\textfont}



\newcount\figcount
\figcount=1


\def\\{\hfill\break}

\newbox\TestBox
\newdimen\setwd
\newskip\belowcaptionskip
\belowcaptionskip=6pt plus 1pt

\def\endinsert{\egroup 
  \if@mid \dimen@\ht\z@ \advance\dimen@\dp\z@
    \advance\dimen@12\p@ \advance\dimen@\pagetotal
    \ifdim\dimen@>\pagegoal\@midfalse\p@gefalse\fi\fi
   \if@mid\vskip\belowcaptionskip\box\z@\par \penalty-
200\vskip\belowcaptionskip
  \else\insert\topins{\penalty100 
    \splittopskip\z@skip
    \splitmaxdepth\maxdimen \floatingpenalty\z@
    \ifp@ge \dimen@\dp\z@
    \vbox to\vsize{\unvbox\z@\kern-\dimen@}
    \else \box\z@\nobreak\vskip\belowcaptionskip\fi}\fi\endgroup}

\def\fig#1#2#3{%
       \setbox\TestBox=\hbox{\tenpoint #3.}\setwd=\wd\TestBox
	\topinsert
       \vskip #1
                \vskip 12pt
        \ifdim\setwd > 23pc
          {\tenit{\smc #2.}\ \ #3}
        \else
         \centerline{\tenit\noindent
                {\smc #2.}\ \  #3}\fi%
	\endinsert}


\newdimen\refindent@
\newdimen\refhangindent@
\newbox\refbox@
\setbox\refbox@=\hbox{\tenrm\baselineskip=11pt [00]}
\refindent@=\wd\refbox@

\def\resetrefindent#1{%
	\setbox\refbox@=\hbox{\tenrm\baselineskip=11pt [#1]}%
	\refindent@=\wd\refbox@}

\def\Refs{%
	\unskip\vskip1pc
	\leftline{\noindent\headfont References}%
	\penalty10000
	\vskip4pt
	\penalty10000
	\refhangindent@=\refindent@
	\global\advance\refhangindent@ by .5em
        \global\everypar{\hangindent\refhangindent@}%
	\parindent=0pt\baselineskip=12pt\tenrm}

\def\sameauthor{\leavevmode\vbox to 1ex{\vskip 0pt plus 100pt
    \hbox to 2em{\leaders\hrule\hfil}\vskip 0pt plus 300pt}}

\def\ref#1\\#2\endref{\leavevmode\hbox to \refindent@{\hfil[#1]}\enspace
#2\par}


\def\rightheadline{\hfill\tensmc\rightrh\hskip2pc\tenrm\folio}
\def\leftheadline{\tenrm\folio\hskip2pc\tensmc\leftrh\hfill}

\global\footline={\hss\tenrm\folio\hss}

\output{\plainoutput}
\def\plainoutput{\shipout\vbox{\makeheadline\pagebody\makefootline}%
  \advancepageno
  \ifnum\pageno>1
	\global\footline={\hfill}%
  \fi
  \ifodd\pageno
	\global\headline={\rightheadline}%
  \else
	\global\headline={\leftheadline}%
  \fi
  \ifnum\outputpenalty>-\@MM \else\dosupereject\fi}
\def\pagebody{\vbox to\vsize{\boxmaxdepth\maxdepth \pagecontents}}
\def\makeheadline{\vbox to\z@{\vskip-22.5\p@
  \line{\vbox to8.5\p@{}\rheadfont\the\headline}\vss}%
    \nointerlineskip}
\def\makefootline{\baselineskip24\p@\vskip-6\p@\line{\the\footline}}
\def\dosupereject{\ifnum\insertpenalties>\z@ 
over
  \line{}\kern-\topskip\nobreak\vfill\supereject\fi}

\def\footnoterule{\vskip11pt\kern -4\p@\hrule width 3pc \kern 3.6\p@ }

%
\catcode`\@=13


\startchapter
\def\leftrh{C. G. Torre}
\def\rightrh{Problems of Time and Observables}
\def\H{{\cal H}}
\title The Problems of Time and Observables:\endtitle
\title Some Recent Mathematical Results\endtitle
\author C. G. Torre\footnote{$^{\dag}$}{Department of Physics, Utah
State University, Logan, UT 84322-4415 USA}\endauthor

Einstein's theory of gravitation can be viewed as a constrained Hamiltonian
system [1].  The constraints are ``first class'', which indicates they generate
``gauge transformations'' on the phase space.  In such systems it is
important to separate cleanly the ``pure gauge'' content of the phase
space from the ``gauge invariant'' information.  In general relativity this
leads,
respectively, to the problem of time and the problem of observables.
Progress on these problems represents an improvement in our
understanding of the
classical dynamics of the gravitational field, but these problems are
particularly relevant to the construction of a quantum theory of gravity.
The problem of
time has been reviewed quite nicely by Kucha\v r [2].  The problem of
observables has been extensively discussed in the literature, but it is hard to
improve on the presentation due to Bergmann [3].  Both of these problems
have been studied from a variety of viewpoints, but rigorous results have
been obtained primarily in the context of simplified model systems.  Here
we give a couple of results on these problems in the context of the full
vacuum theory.

The spacetime manifold $M$ is assumed to have the topology
$M=R\times\Sigma$, where $\Sigma$ is a compact 3-manifold.  We will
consider the phase space $\Gamma$ for gravitational dynamics to
be a cotangent bundle over the space of Riemannian metrics on $\Sigma$.
Dynamics takes place on the constraint
surface $\overline\Gamma$ defined by the Hamiltonian and momentum
constraints ${\cal H}=0$ and ${\cal H}_a=0$.

Our first result deals with the problem of time; specifically, we examine a
proposal of Kucha\v r to solve the problem of time by identifying
$\Gamma$ with the phase space of some ``parametrized field theory'' [2],
[4].  The
idea is to isolate four fields $X^\alpha$ from $\Gamma$ that can, at least
on $\overline\Gamma$,
be viewed as representing a spacelike embedding $X^\alpha\colon\Sigma\to
M$.
The constraints $\H=0=\H_a$ are then interpreted as conditions that
identify the momenta
$P_\alpha$ conjugate to the embedding with the energy-momentum
densities of the remaining dynamical degrees of freedom.  In order to
investigate the viability of this scheme, we compare the constraint surface
of
a generic parametrized theory with $\overline\Gamma$.  To this end, let
$T^*{\cal E}$ denote a cotangent bundle over the space of embeddings
$X^\alpha\colon\Sigma\to M$, and let $\Omega$ denote an infinite
dimensional
symplectic manifold. The phase space $\Omega$ is to represent the ``true
degrees of freedom'' of the gravitational field; points of $\Omega$ will be
labeled $Z^A$.    The phase space $\Upsilon$
for a generic parametrized field theory is given by $\Upsilon=T^*{\cal
E}\times\Omega$.  Dynamics takes place on the constraint surface
$\overline\Upsilon$, which is defined as follows.  Let $h_\alpha$ denote
four densities constructed from
the embeddings $X^\alpha$ and dynamical variables $Z^A$.
$\overline\Upsilon$
is then the subspace of $\Upsilon$ for which $P_\alpha+h_\alpha=0$.

To
implement the proposal of Kucha\v r we must find a canonical
transformation (symplectic diffeomorphism)
$\Phi\colon\Upsilon\to\Gamma$ such that
$\Phi(\overline\Upsilon)=\overline\Gamma$.  However, no such
diffeomorphism exists [5].
The proof of this result rests on the fact that $\overline\Upsilon$ is a
manifold, which can be seen by applying the implicit function theorem.  On
the other hand, provided the topology of $\Sigma$ does not prohibit the
existence of vacuum spacetimes with Killing vectors, the constraint surface
$\overline\Gamma$ is not a manifold [6].  Hence there can be no bijection
identifying $\overline\Upsilon$ and $\overline\Gamma$.  The essence of
the difficulty with making this identification is that, in a spacetime with
an isometry, it is impossible to distinguish embeddings that differ by an
action of the isometry using only the intrinsic and extrinsic geometry of the
embedded
hypersurface.  There are probably ways to mitigate this difficulty but,
strictly speaking, general
relativity is not a parametrized field theory.

Our second result concerns the existence of gravitational observables
that can be expressed as spatial integrals of densities which are built locally
from the
canonical variables and their derivatives to any order.  Recall that the
constraint functions $(\H,\H_a)$, being ``first class'', generate 1-parameter
families of canonical
transformations that leave $\overline\Gamma$ invariant.  An observable is
defined as a function $F\colon\Gamma\to R$ whose restriction to
$\overline\Gamma$ is invariant under the flow generated by the
constraints.  What this means in practice is that one looks for functions on
$\Gamma$ whose Poisson brackets with $\H$ and $\H_a$ vanish on
$\overline\Gamma$.  Any observable that vanishes on the constraint
surface is called {\it trivial}.  Any two observables that differ by a trivial
observable will carry the same physical information, therefore we identify
any two observables if their difference vanishes on $\overline\Gamma$.
Next, recall that there are an infinite number of Hamiltonians that
generate vacuum spacetimes from initial data on $\overline\Gamma$, but
they are all linear combinations of the constraint functions.  More
precisely, the Einstein Hamiltonian $H$ is of the form $H=\int_\Sigma(N\H
+ N^a\H_a)$, where $N$ is a (positive) function
on $\Sigma$, and $N^a$ is a vector field on $\Sigma$.  We then see
that the observables, as defined above, are constants of the motion.  The
problem of observables amounts to finding enough constants of the motion
to uniquely specify any vacuum spacetime (up to
diffeomorphisms).  Unfortunately, not a single observable is known.
Indeed, aside from the work of Kucha\v r [7], very little is known about
the
form such observables can take.

Let $(q_{ab},p^{ab})\in\Gamma$, {\it i.e.}, $q_{ab}$ is a Riemannian
metric on $\Sigma$ and $p^{ab}$ is a symmetric tensor density on
$\Sigma$.
We will classify all {\it local observables}, which are defined to be
constants of motion that can be expressed as integrals on $\Sigma$ of
densities built
locally from the phase space variables $(q_{ab},p^{ab})$ and their
derivatives to any order.  For
example, the Hamiltonian $H$ is a local observable albeit a trivial one, {\it
i.e.}, it is equivalent to zero.  Observables generate 1-parameter families of
canonical transformations that
preserve $\overline\Gamma$.  Infinitesimally, the local observables
generate transformations of $\overline\Gamma$ that are built locally from
the canonical variables and their derivatives.   It is not too hard to see
that,
in spacetime language, the existence of a local observable corresponds to
the existence of an infinitesimal transformation of the {\it spacetime}
metric that (i) is built locally from the spacetime metric and its derivatives,
(ii) maps a solution of the Einstein equations to a nearby solution.  Such
transformations are called ``generalized symmetries'' by applied
mathematicians.  Recently all generalized symmetries of the vacuum
Einstein equations have been classified [9].  They consist of a trivial scaling
symmetry and the familiar diffeomorphism symmetry.  The former cannot
be implemented as a symplectic map of $\Gamma$, while the latter is
generated by the constraint functions themselves.  Because there are no
other symmetries, there can be no observables (save the trivial constraints)
built as local functionals of the canonical variables [10].

To summarize, we have ruled out the simplest putative resolutions of the
problems of time and observables.  We cannot use parametrized field
theory to solve the problem of time because, strictly speaking, general
relativity is not a parametrized field theory.  And we have seen that there
are essentially no
local observables for vacuum spacetimes.

\Refs

\ref 1\\ A. Hanson, T. Regge, C. Teitelboim, {\it Constrained Hamiltonian
Systems}, Academia Nazionale Dei Lincei, Rome, 1976;  A. Ashtekar, {\it
Lectures on Non-Perturbative
Canonical Gravity}, World Scientific, Singapore, 1991.
\endref

\ref 2\\K. Kucha\v r, ``Time and Interpretations of
Quantum Gravity'', in the proceedings of {\it The Fourth Canadian
Conference on
General Relativity and Relativistic Astrophysics}, edited by G. Kunstatter,
D. Vincent, and J. Williams, World Scientific, Singapore, 1992.
\endref

\ref 3\\P. Bergmann, {\it Rev. Mod. Phys.} {\bf 33}, 510 (1961).
\endref

\ref 4\\K. Kucha\v r, {\it J. Math. Phys.} {\bf 13}, 768 (1972).
\endref

\ref 5\\C. Torre, {\it Phys. Rev.} D{\bf 46}, R3231 (1992).
\endref

\ref 6\\A. Fischer, J. Marsden, and V. Moncrief, {\it Ann. Inst. H.
Poincar\'e} {\bf 33}, 147 (1980).
\endref

\ref 7\\K. Kucha\v r, {\it J. Math. Phys.} {\bf 22}, 2640 (1981).
\endref

\ref 8\\P. Olver, {\it Applications of Lie Groups to Differential
Equations}, Springer-Verlag, New York, 1993.
\endref

\ref 9\\C. Torre and I. Anderson, {\it Phys. Rev. Lett} {\bf 70}, 3525
(1993); {\it Generalized Symmetry Analysis of the Vacuum Einstein
Equations}, Utah State University Preprint, 1994; {\it Two Component
Spinors and Natural Coordinates for the Prolonged Einstein Equation
Manifolds}, Utah State University Technical Report, 1994.
\endref

\ref 10\\C. Torre, {\it Phys. Rev.} D{\bf 48}, R2373 (1993).
\endref

\bye